\documentclass[12pt, draftcls, onecolumn]{IEEEtran_WCM_12_00220}
\usepackage{setspace}
\usepackage{multirow}
\usepackage[dvips]{graphicx}
\usepackage[cmex10]{amsmath}
\usepackage{amsmath}
\usepackage{amssymb}
\usepackage{amsfonts}
\DeclareGraphicsExtensions{ps,eps,pst,pdf}
\usepackage{cite}

\begin{document}

\title{Wideband Spectrum Sensing for Cognitive Radio Networks: A Survey}

\vspace{-2em}

\author{Hongjian~Sun and Arumugam~Nallanathan, King's College London\\
Cheng-Xiang~Wang, Heriot-Watt University\\
Yunfei~Chen, University of Warwick
\thanks{Copyright $\copyright$ 2013 IEEE. Personal use of this material is permitted. However, permission to use this material for any other purposes must be obtained from the IEEE by sending a request to pubs-permissions@ieee.org.}
\thanks{This paper has been accepted to be published in IEEE Wireless Communications, 2013. }}

\maketitle

\begin{abstract}
Cognitive radio has emerged as one of the most promising candidate solutions to improve spectrum utilization in next generation cellular networks. A crucial requirement for future cognitive radio networks is wideband spectrum sensing: secondary users reliably detect spectral opportunities across a wide frequency range. In this article, various wideband spectrum sensing algorithms are presented, together with a discussion of the pros and cons of each algorithm and the challenging issues. Special attention is paid to the use of sub-Nyquist techniques, including compressive sensing and multi-channel sub-Nyquist sampling techniques.

\end{abstract}


\newpage

\section{Introduction}
\label{section0}

Radio frequency (RF) spectrum is a valuable but tightly regulated resource due to its unique and important role in wireless communications. With the proliferation of wireless services, the demands for the RF spectrum are constantly increasing, leading to scarce spectrum resources. On the other hand, it has been reported that localized temporal and geographic spectrum utilization is extremely low \cite{nsf}. Currently, new spectrum policies are being developed by the Federal Communications Commission (FCC) that will allow secondary users to opportunistically access a licensed band, when the primary user (PU) is absent. Cognitive radio \cite{Wan09a, Wan09b} has become a promising solution to solve the spectrum scarcity problem in the next generation cellular networks by exploiting opportunities in time, frequency, and space domains.

Cognitive radio is an advanced software-defined radio that automatically detects its surrounding RF stimuli and intelligently adapts its operating parameters to network infrastructure while meeting user demands. Since cognitive radios are considered as secondary users for using the licensed spectrum, a crucial requirement of cognitive radio networks is that they must efficiently exploit under-utilized spectrum (denoted as spectral opportunities) without causing harmful interference to the PUs. Furthermore, PUs have no obligation to share and change their operating parameters for sharing spectrum with cognitive radio networks. Hence, cognitive radios should be able to independently detect spectral opportunities without any assistance from PUs; this ability is called spectrum sensing, which is considered as one of the most critical components in cognitive radio networks.

Many narrowband spectrum sensing algorithms have been studied in the literature \cite{survey} and references therein, including matched-filtering, energy detection \cite{ed}, and cyclostationary feature detection. While present narrowband spectrum sensing algorithms have focused on exploiting spectral opportunities over narrow frequency range, cognitive radio networks will eventually be required to exploit spectral opportunities over wide frequency range from hundreds of megahertz (MHz) to several gigahertz (GHz) for achieving higher opportunistic throughput. This is driven by the famous Shannon's formula that, under certain conditions, the maximum theoretically achievable bit rate is directly proportional to the spectral bandwidth.  Hence, different from narrowband spectrum sensing, wideband spectrum sensing aims to find more spectral opportunities over wide frequency range and achieve higher opportunistic aggregate throughput in cognitive radio networks. However, conventional wideband spectrum sensing techniques based on standard analog-to-digital converter (ADC) could lead to unaffordably high sampling rate or implementation complexity; thus, revolutionary wideband spectrum sensing techniques become increasingly important.

In the remainder of this article, we first briefly introduce the traditional spectrum sensing algorithms for narrowband sensing in Section~\ref{section2}. Some challenges for realizing wideband spectrum sensing are then discussed in Section~\ref{section3}. In addition, we categorize the existing wideband spectrum sensing algorithms based on their implementation types, and review the state-of-the-art techniques for each category. Future research challenges for implementing wideband spectrum sensing are subsequently identified in Section~\ref{section4}, after which  concluding remarks are given in Section~\ref{section5}.

\section{Narrowband Spectrum Sensing}
\label{section2}

\begin{figure}[!ht]
\centerline{\includegraphics[width=5.5in]{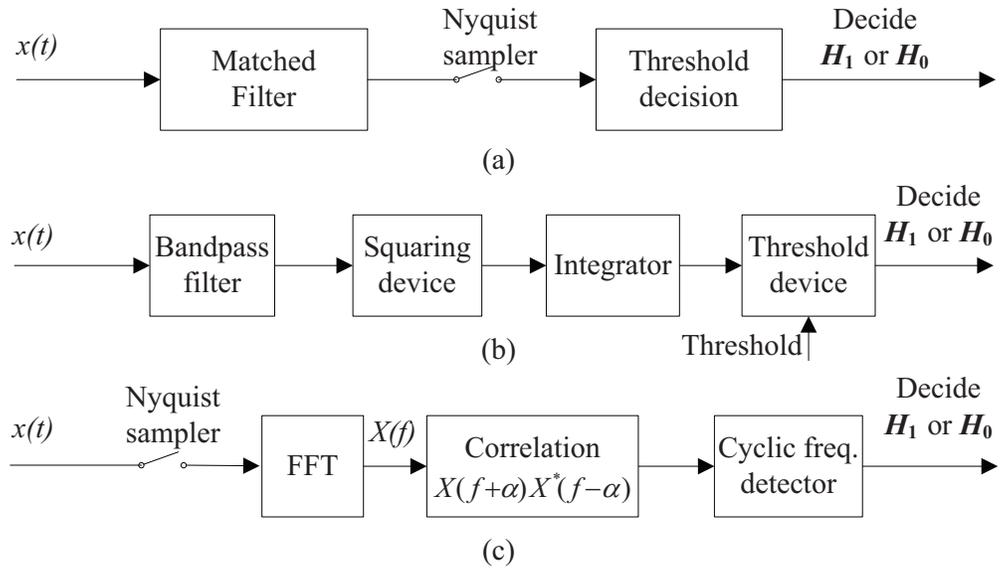}}
\caption{Block diagrams for narrowband spectrum sensing algorithms: (a) matched-filtering, (b)  energy detection, and (c) cyclostationary feature detection.}
\label{fig1}
\end{figure}

The most efficient way to sense spectral opportunities is to detect active primary transceivers in the vicinity of cognitive radios. However, as primary receivers may be passive, such as TVs, some receivers are difficult to detect in practice. An alternative is to detect the primary transmitters by using traditional narrowband sensing algorithms, including matched-filtering, energy detection, and cyclostationary feature detection as shown in Fig.~\ref{fig1}. Here, the term ``narrowband'' implies that the frequency range is sufficiently narrow such that the channel frequency response can be considered flat. In other words, the bandwidth of our interest is less than the coherence bandwidth of the channel. The implementation of these narrowband algorithms requires different conditions, and their detection performance are correspondingly distinguished. The advantages and disadvantages of these algorithms are summarized in Table~\ref{table:compare}.

The matched-filtering method is an optimal approach for spectrum sensing since it maximizes the signal-to-noise ratio (SNR) in the presence of additive noise. This advantage is achieved by correlating the received signal with a template for detecting the presence of a known signal in the received signal. However, it relies on prior knowledge of the PUs and requires cognitive radios to be equipped with carrier synchronization and timing devices, leading to increased implementation complexity.
Energy detection \cite{ed} is a non-coherent detection method that avoids the need for prior knowledge of the PUs and the complicated receivers required by a matched filter. Both the implementation and the computational complexity are relatively low. A major drawback is that it has poor detection performance under low SNR scenarios and cannot differentiate between the signals from PUs and the interference from other cognitive radios.
Cyclostationary feature detection method detects and distinguishes between different types of primary signals by exploiting their cyclostationary features. However, the computational cost of such an approach is relatively high, because it requires to calculate a two-dimensional function dependent on both frequency and cyclic frequency.

\section{Wideband Spectrum Sensing}
\label{section3}

Against narrowband techniques as mentioned above, wideband spectrum sensing techniques aim to sense a frequency bandwidth that exceeds the coherence bandwidth of the channel. For example, for exploiting spectral opportunities in the whole ultra-high frequency (UHF) TV band (between 300 MHz and 3 GHz), wideband spectrum sensing techniques should be employed. We note that narrowband sensing techniques cannot be directly used for performing wideband spectrum sensing, because they make a single binary decision for the whole spectrum and thus cannot identify individual spectral opportunities that lie within the wideband spectrum. As shown in Table~\ref{table1}, wideband spectrum sensing can be broadly categorized into two types: Nyquist wideband sensing and sub-Nyquist wideband sensing. The former type processes digital signals taken at or above the Nyquist rate, whereas the latter type acquires signals using sampling rate lower than the Nyquist rate. In the rest of this article, we will provide an overview of the state-of-the-art wideband spectrum sensing algorithms and discuss the pros and cons of each algorithm.

\subsection{Nyquist Wideband Sensing}
\label{3.1}

\begin{figure}[!ht]
\centerline{\includegraphics[width=6.3in]{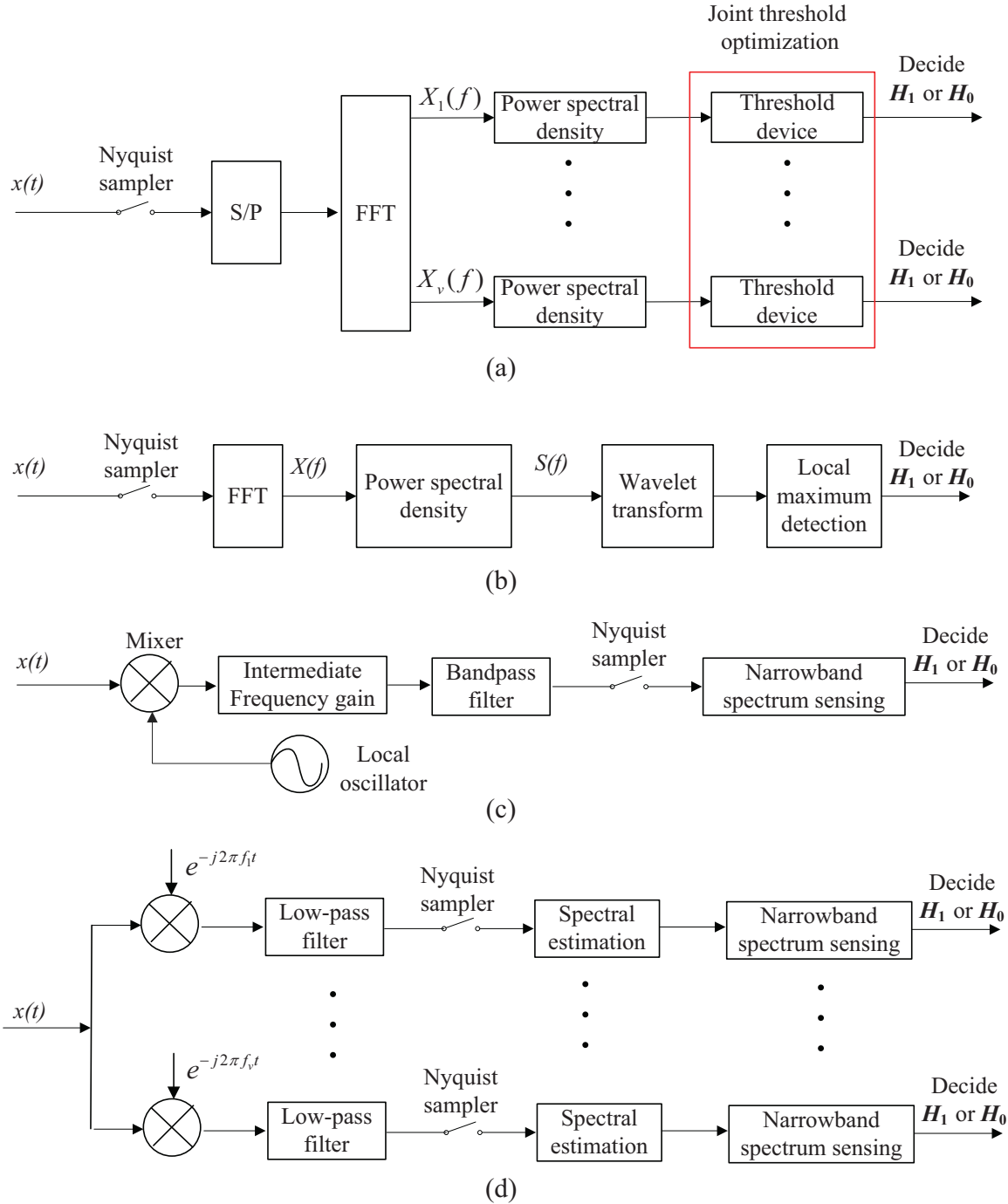}}
\caption{Block diagrams for Nyquist wideband sensing algorithms: (a) Multiband joint detection, (b) Wavelet detection, (c) Sweep-tune detection, and (d) Filter-bank detection.}
\label{fig2}
\end{figure}

A simple approach of wideband spectrum sensing is to directly acquire the wideband signal using a standard ADC and then use digital signal processing techniques to detect spectral opportunities. For example, Quan {\em et al.} \cite{quan} proposed a multi-band joint detection algorithm that can sense the primary signal over multiple frequency bands. As shown in Fig.~\ref{fig2}(a), the wideband signal $x(t)$ was firstly sampled by a high sampling rate ADC, after which a serial to parallel conversion circuit (S/P) was used to divide sampled data into parallel data streams. Fast Fourier transform (FFT) was used to convert the wideband signals to the frequency domain. The wideband spectrum $X(f)$ was then divided into a series of narrowband spectra $X_1(f), \cdots, X_v(f)$. Finally, spectral opportunities were detected using binary hypotheses tests, where $\mathcal{H}_{0}$ denotes the absence of PUs and $\mathcal{H}_{1}$ denotes the presence of PUs. The optimal detection threshold was jointly chosen by using optimization techniques. Such an algorithm can achieve better performance than the single band sensing case.

Furthermore, by also using a standard ADC, Tian and Giannakis proposed a wavelet-based spectrum sensing algorithm in \cite{Tian2006}. In this algorithm, the power spectral density (PSD) of the wideband spectrum (denoted as $S(f)$) was modeled as a train of consecutive frequency subbands, where the PSD is smooth within each subband but exhibits discontinuities and irregularities on the border of two neighboring subbands. The wavelet transform was then used to locate the singularities of the wideband PSD, and the wideband spectrum sensing was formulated as a spectral edge detection problem as shown in Fig.~\ref{fig2}(b).

However, special attention should be paid to the signal sampling procedure. In these algorithms, sampling signals should follow Shannon's celebrated
theorem: the sampling rate must be at least twice the maximum frequency present in the signal (known as Nyquist rate) in order to avoid spectral aliasing. Suppose that the wideband signal has frequency range $0\sim 10$ GHz, it should be uniformly sampled by a standard ADC at or above the Nyquist rate $20$ GHz which will be unaffordable for next generation cellular networks. Therefore, sensing wideband spectrum presents significant challenges on building sampling hardware that operates at a sufficiently high rate, and designing high-speed signal processing algorithms. With current hardware technologies, high-rate ADCs with high resolution and reasonable power consumption (e.g., 20 GHz sampling rate with 16 bits resolution) are difficult to implement. Even if it comes true, the real-time digital signal processing of sampled data could be very expensive.

One naive approach that could relax the high sampling rate requirement is to use superheterodyne (frequency mixing) techniques that ``sweep'' across the frequency range of interest as shown in Fig.~\ref{fig2}(c). A local oscillator (LO) produces a sine wave that mixes with the wideband signal and down-converts it to a lower frequency. The down-converted signal is then filtered by a band-pass filter (BPF), after which existing narrowband spectrum sensing techniques in Section~\ref{section2} can be applied. This sweep-tune approach can be realized by using either a tunable BPF or a tunable LO. However, this approach is often slow and inflexible due to the sweep-tune operation.

Another solution would be the filter bank algorithm presented by Farhang-Boroujeny \cite{bank} as shown in Fig.~\ref{fig2}(d). A bank of prototype filters (with different shifted central frequencies) was used to process the wideband signal. The base-band can be directly estimated by using a prototype filter, and other bands can be obtained through modulating the prototype filter. In each band, the corresponding portion of the spectrum for the wideband signal was down-converted to base-band and then low-pass filtered. This algorithm can therefore capture the dynamic nature of wideband spectrum by using low sampling rates. Unfortunately, due to the parallel structure of the filter bank, the implementation of this algorithm requires a large number of RF components.

\subsection{Sub-Nyquist Wideband Sensing}
\label{3.2}

Due to the drawbacks of high sampling rate or high implementation complexity in Nyquist systems, sub-Nyquist approaches are drawing more and more attention in both academia and industry. Sub-Nyquist wideband sensing refers to the procedure of acquiring wideband signals using sampling rates lower than the Nyquist rate and detecting spectral opportunities using these partial measurements. Two important types of sub-Nyquist wideband sensing are compressive sensing-based wideband sensing and multi-channel sub-Nyquist wideband sensing. In the subsequent paragraphs, we give some discussions and comparisons regarding these sub-Nyquist wideband sensing algorithms.

\subsubsection{Compressive Sensing-based Wideband Sensing}
\label{3.2.1}

Compressive sensing is a technique that can efficiently acquire a signal using relatively few measurements, by which unique representation of the signal can be found based on the signal's sparseness or compressibility in some domain. As the wideband spectrum is inherently sparse due to its low spectrum utilization, compressive sensing becomes a promising candidate to realize wideband spectrum sensing by using sub-Nyquist sampling rates.
Tian and Giannakis firstly introduced compressive sensing theory to sense wideband spectrum in \cite{scs}. This technique used fewer samples closer to the information rate, rather than the inverse of the bandwidth, to perform wideband spectrum sensing. After reconstruction of the wideband spectrum, wavelet-based edge detection was used to detect spectral opportunities across wideband spectrum.

Furthermore, to improve the robustness against noise uncertainty, Tian {\em et al.} \cite{cyclic} studied a cyclic feature detection-based compressive sensing algorithm for wideband spectrum sensing. It can successfully extract second-order statistics of wideband signals from digital samples taken at sub-Nyquist rates. The 2-D cyclic spectrum (spectral correlation function) of a wideband signal can be directly reconstructed from the compressive measurements. In addition, such an algorithm is also valid for reconstructing the power spectrum of wideband signal, which is useful if the energy detection algorithm is used for detecting spectral opportunities.

For further reducing the data acquisition cost, Zeng {\em et al.} \cite{distributed} proposed a distributed compressive sensing-based wideband sensing algorithm for cooperative multi-hop cognitive radio networks. By enforcing consensus among local spectral estimates, such a collaborative approach can benefit from spatial diversity to mitigate the effects of wireless fading. In addition, decentralized consensus optimization algorithm was proposed that aims to achieve high sensing performance at a reasonable computational cost.

However, compressive sensing has concentrated on finite-length and discrete-time signals. Thus, innovative technologies are required to extend the compressive sensing to continuous-time signal acquisition, i.e., implementing compressive sensing in analog domain.
To realize the analog compressive sensing, Tropp {\em et al.} \cite{beyond} proposed an analog-to-information converter (AIC), which could be a good basis for the above-mentioned algorithms. As shown in Fig.~\ref{fig3}(a), the AIC-based model consists of a pseudo-random number generator, a mixer, an accumulator, and a low-rate sampler. The pseudo-random number generator produces a discrete-time sequence that demodulates the signal $x(t)$ by a mixer. The accumulator is used to sum the demodulated signal for $1/w$ seconds, while its output signal is sampled using a low sampling rate. After that, the sparse signal can be directly reconstructed from partial measurements using compressive sensing algorithms. Unfortunately, it has been identified that the performance of AIC model can be easily affected by design imperfections or model mismatches.

\subsubsection{Multi-channel Sub-Nyquist Wideband Sensing}
\label{3.2.2}

\begin{figure}[!ht]
\centerline{\includegraphics[width=5in]{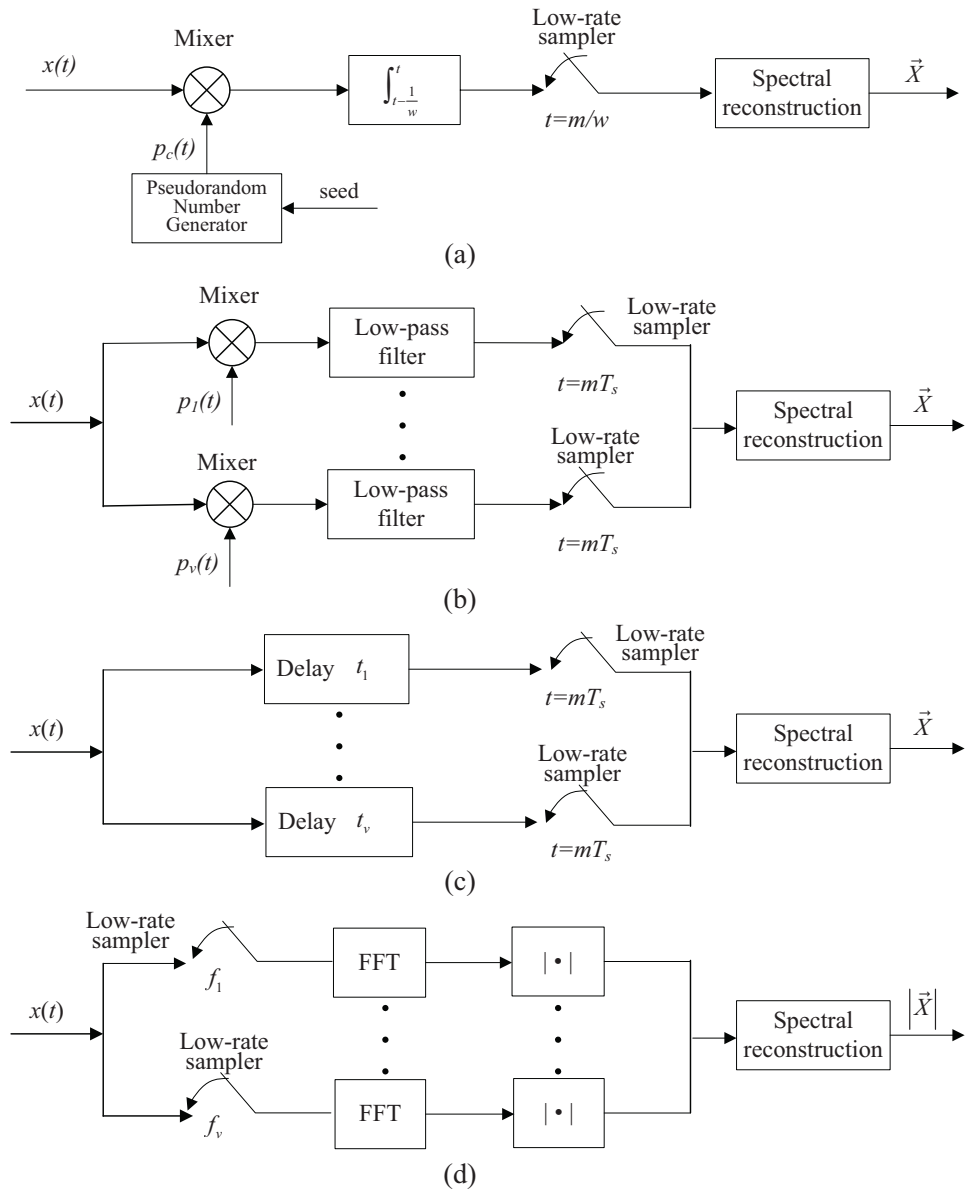}}
\caption{Block diagrams for sub-Nyquist wideband sensing algorithms: (a) Analog-to-information converter-based wideband sensing, (b) Modulated wideband converter-based wideband sensing, (c) Multi-coset sampling-based wideband sensing, and (d) Multi-rate sub-Nyquist sampling-based wideband sensing.}
\label{fig3}
\end{figure}

To circumvent model mismatches, Mishali and Eldar proposed a modulated wideband converter (MWC) model in \cite{was1} by modifying the AIC model. The main difference between MWC and AIC is that MWC has multiple sampling channels, with the accumulator in each channel replaced by a general low-pass filter. One significant benefit of introducing parallel channel structure in Fig.~\ref{fig3}(b) is that it provides robustness against the noise and model mismatches. In addition, the dimension of the measurement matrix is reduced, making the spectral reconstruction more computationally efficient.

An alternative multi-channel sub-Nyquist sampling approach is the multi-coset sampling as shown in Fig.~\ref{fig3}(c). The multi-coset sampling is equivalent to choosing some samples from a uniform grid, which can be obtained using a sampling rate $f_s$ higher than the Nyquist rate. The uniform grid is then divided into blocks of $m$ consecutive samples, and in each block $v (v<m)$ samples are retained while the rest of samples are skipped. Thus, the multi-coset sampling is often implemented by using $v$ sampling channels with sampling rate of $\frac{f_s}{m}$, with different sampling channels having different time offsets.  To obtain a unique solution for the wideband spectrum from these partial measurements, the sampling pattern should be carefully designed. In \cite{unique1}, some sampling patterns were proved to be valid for unique signal reconstruction. The advantage of multi-coset approach is that the sampling rate in each channel is $m$ times lower than the Nyquist rate. Moreover, the number of measurements is only $v$-$m$th of that in the Nyquist sampling case. One drawback of the multi-coset approach is that the channel synchronization should be met such that accurate time offsets between sampling channels are required to satisfy a specific sampling pattern for a robust spectral reconstruction.

To relax the multi-channel synchronization requirement, asynchronous multi-rate wideband sensing approach was studied in \cite{hongjian}. In this approach, sub-Nyquist sampling was induced in each sampling channel to wrap the sparse spectrum occupancy map onto itself; the sampling rate can therefore be significantly reduced. By using different sampling rates in different sampling channels as shown in Fig.~\ref{fig3}(d), the performance of wideband spectrum sensing can be improved. Specifically, in the same observation time, the numbers of samples in multiple sampling channels are chosen as different consecutive prime numbers. Furthermore, as only the magnitudes of sub-Nyquist spectra are of interest, such a multi-rate wideband sensing approach does not require perfect synchronization between multiple sampling channels, leading to easier implementation.

\section{Open Research Challenges}
\label{section4}

In this section, we identify the following research challenges that need to be addressed for implementing a feasible wideband spectrum sensing device for future cognitive radio networks.

\subsection{Sparse Basis Selection}

Nearly all sub-Nyquist wideband sensing techniques require that the wideband signal should be sparse in a suitable basis. Given the low spectrum utilization, most of existing wideband sensing techniques assumed that the wideband signal is sparse in the frequency domain, i.e., the sparsity basis is a Fourier matrix. However, as the spectrum utilization improves, e.g., due to the use of cognitive radio techniques in future cellular networks, the wideband signal may not be sparse in the frequency domain any more.  Thus, a significant challenge in future cognitive radio networks is how to perform wideband sensing using partial measurements, if the wideband signal is not sparse in the frequency domain. It will be essential to study appropriate wideband sensing techniques that are capable of exploiting sparsity in any known sparsity basis. Furthermore, in practice, it may be difficult to acquire sufficient knowledge about the sparsity basis in cognitive radio networks, e.g., when we cannot obtain enough prior knowledge about the primary signals. Hence, future cognitive radio networks will be required to perform wideband sensing when the sparsity basis is not known. In this context, a challenging issue is to study ``blind'' sub-Nyquist wideband sensing algorithms, where we do not require prior knowledge about the sparsity basis for the sub-Nyquist sampling or the spectral reconstruction.

\subsection{Adaptive Wideband Sensing}

\begin{figure}[!ht]
\centerline{\includegraphics[width=5.5in]{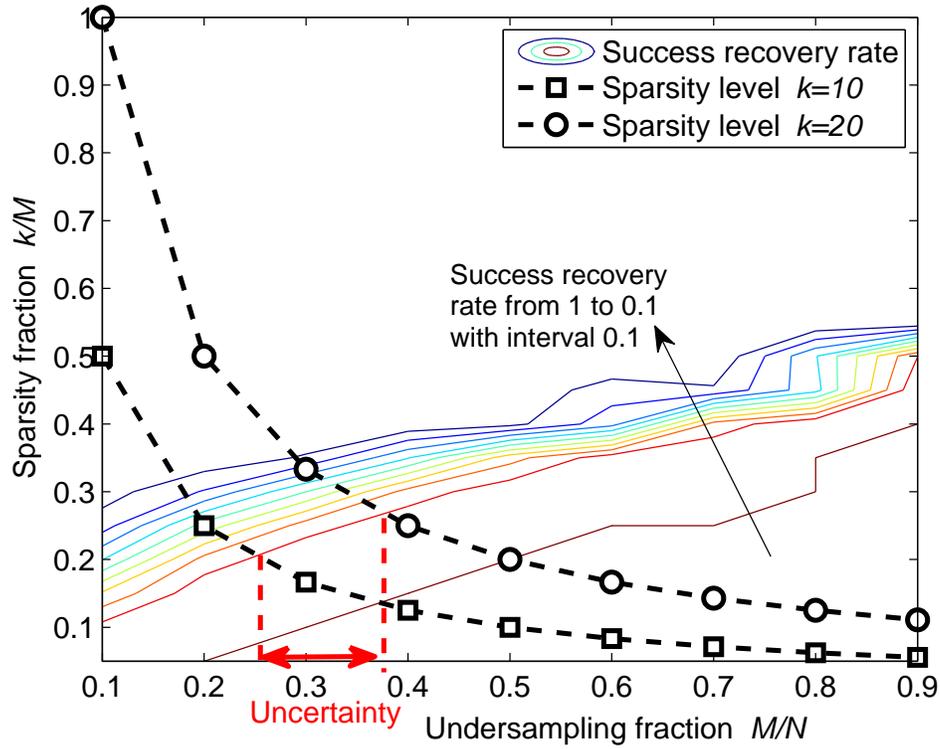}}
\vspace{-1em}
\caption{An example of a sub-Nyquist system, where the sparsity level uncertainty will result in more number of measurements for a fixed successful recovery rate. In simulations, assuming the number of measurements under the Nyquist rate $N=200$, we varied the number of measurements $M$ from 20 to 180 in eight equal-length steps. The sparsity level $k$ was set to between 1 and $M$. The measurement matrix was assumed to be Gaussian. The figure was obtained with 5000 trials of each parameter setting. }
\label{fig4}
\end{figure}

In most of sub-Nyquist wideband sensing systems, the required number of measurements will proportionally change when the sparsity level of wideband signal varies. Therefore, sparsity level estimation is often required for choosing an appropriate number of measurements in cognitive radio networks. However, in practice, the sparsity level of wideband signal is often time-varying and difficult to estimate, because of either the dynamic activities of PUs or the time-varying fading channels between PUs and cognitive radios. Due to this sparsity level uncertainty, most of sub-Nyquist wideband sensing systems should pessimistically choose the number of measurements, leading to more energy consumption in cellular networks. As shown in Fig.~\ref{fig4}, more measurements (i.e., $0.38N$ rather than $0.25N$ measurements for achieving the success recovery rate 0.9) are required for the sparsity uncertainty between 10 and 20, which does not fully exploit the advantages of using sub-Nyquist sampling technologies. Hence, future cognitive radio networks should be capable of performing wideband sensing, given the unknown or time-varying sparsity level. In such a scenario, it is very challenging to develop adaptive wideband sensing techniques that can intelligently/quickly choose an appropriate number of compressive measurements without the prior knowledge of the sparsity level.

\subsection{Cooperative Wideband Sensing}

In a multipath or shadow fading environment, the primary signal as received at cognitive radios may be severely degraded, leading to unreliable wideband sensing results in each cognitive radio. In this situation, future cognitive radio networks should employ cooperative strategies for improving the reliability of wideband sensing by exploiting spatial diversity. Actually, in a cluster-based cognitive radio network, the wideband spectra as observed by different cognitive radios could share some common spectral components, while each cognitive radio may observe some innovative spectral components. Thus, it is possible to fuse compressive measurements from different nodes and exploit the spectral correlations among cognitive radios in order to save the total number of measurements and thus the energy consumption in cellular networks. Such a data fusion-based cooperative technique, however, will lead to heavy data transmission burden in the common control channels. It is therefore challenging to develop data fusion-based cooperative wideband sensing techniques subject to relaxed data transmission burden. An alternative is to develop decision fusion-based wideband sensing techniques, if each cognitive radio is able to detect wideband spectrum independently. Due to the limited computational resource in cellular networks, the challenge that remains in the decision fusion-based cooperative approach is how to appropriately combine information in real time.

\section{Concluding Remarks}
\label{section5}

In this article, we first addressed the challenges in the design and implementation of wideband spectrum sensing algorithms for the cognitive radio-based next generation cellular networks. Then, we categorized the existing wideband spectrum sensing algorithms based on their sampling types and discussed the pros and cons of each category. Moreover, motivated by the fact that wideband spectrum sensing is critical for reliably finding spectral opportunities and achieving opportunistic spectrum access for next generation cellular networks, we made a brief survey of the state-of-the-art wideband spectrum sensing algorithms. Finally, we presented several open research issues for implementing wideband spectrum sensing.

\section*{Acknowledgement}

H. Sun and A. Nallanathan acknowledge the support of the UK Engineering and Physical Sciences Research Council (EPSRC) with Grant No. EP/I000054/1. C.-X. Wang acknowledges the support from the RCUK for the UK-China Science Bridges Project: R\&D on (B)4G Wireless Mobile Communications, the Key Laboratory of Cognitive Radio and Information Processing (Guilin University of Electronic Technology), Ministry of Education, China (Grant No.: 2011KF01), the Fundamental Research Program of Shenzhen City (Grant No. JCYJ20120817163755061), and SNCS research center at University of Tabuk under the grant from the Ministry of Higher Education in Saudi Arabia.

\bibliographystyle{IEEEtran}

\begin{IEEEbiography}[{\includegraphics[width=1in,clip,keepaspectratio]{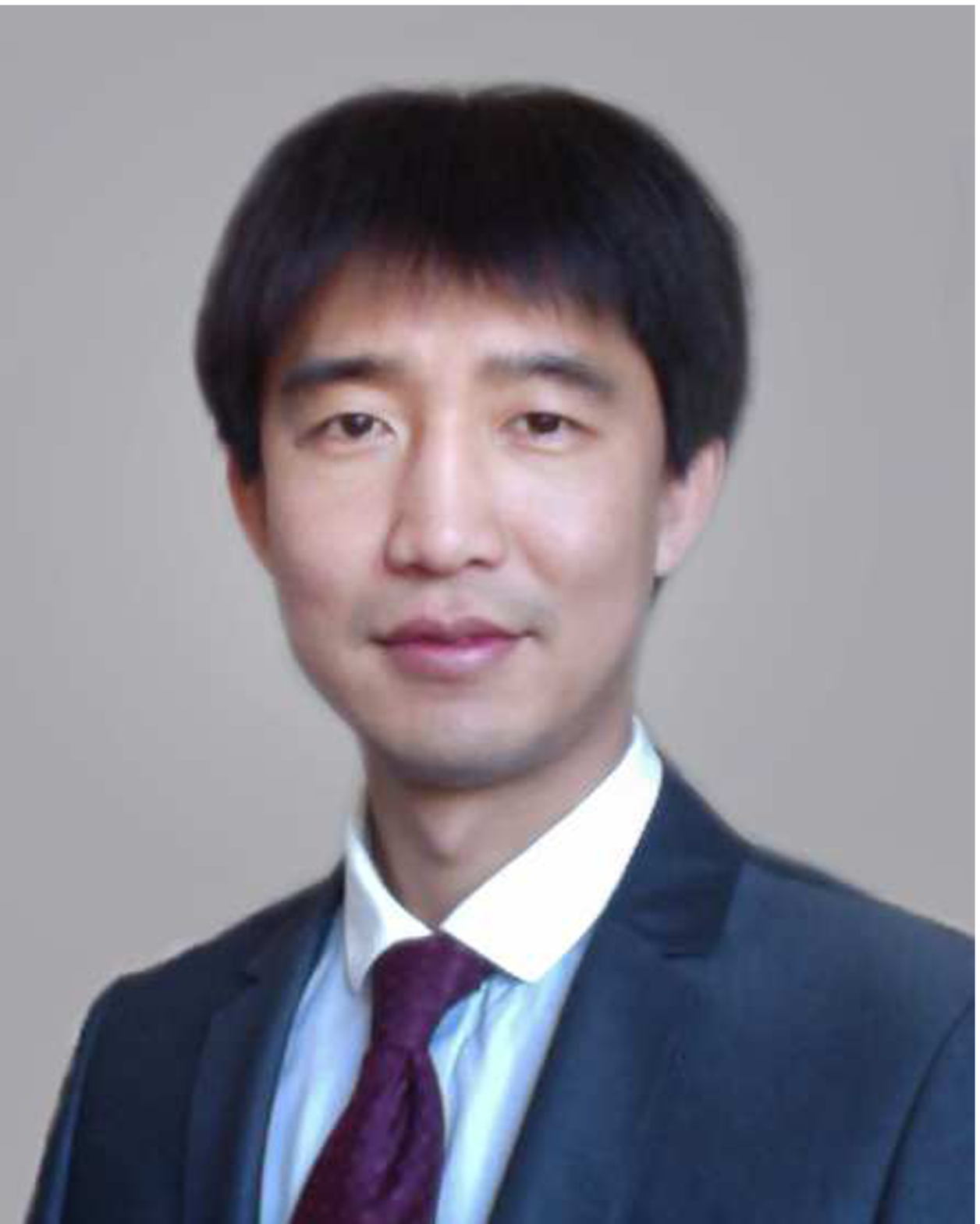}}]{Hongjian Sun} [S'07-M'11]
(hongjian@ieee.org) obtained Ph.D. degree at the University of Edinburgh, UK, where he also received Wolfson Microelectronics Scholarship, in 2010. He then joined King's College London, UK, as a Postdoctoral Research Associate. In 2011-2012, he was a visiting Postdoctoral Research Associate at Princeton University, USA. His recent research interests include Cognitive Radio, Smart Grid, Cooperative Communication, and Compressive Sensing. He has made 1 contribution to the IEEE 1900.6a Standard, and published 1 book chapter and more than 40 papers in refereed journals and conferences.
\end{IEEEbiography}

\begin{IEEEbiography}[{\includegraphics[width=1in,height=1.25in,clip,keepaspectratio]{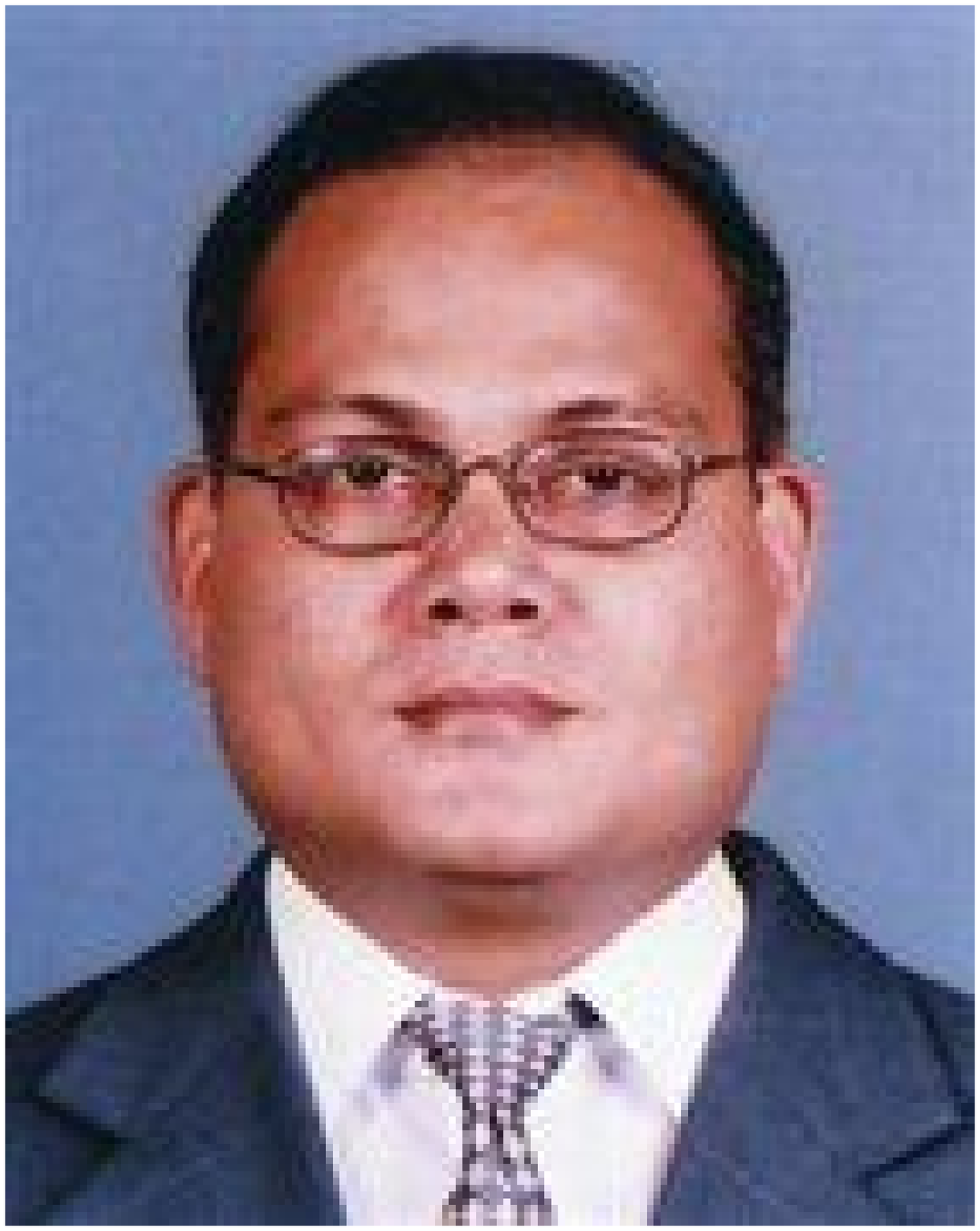}}]{Arumugam Nallanathan}[S'97-M'00-SM'05]
(nallanathan@ieee.org) is currently a Reader in Communications and served as the Head of Graduate Studies in the School of Natural and Mathematical Sciences at King's College London, London, U.K. From 2000 to
2007, he was an Assistant Professor in the Department of Electrical
and Computer Engineering at the National University of Singapore,
Singapore. His research interests include smart grid, cognitive
radio, and relay networks. He has authored nearly 200 journal and
conference papers. He is an Editor for IEEE Transactions on
Communications, IEEE Transactions on Vehicular Technology, IEEE
Wireless Communications Letters and IEEE Signal Processing Letters.
He served as an Editor for IEEE Transactions on Wireless
Communications (2006-2011).
\end{IEEEbiography}

\begin{IEEEbiography}[{\includegraphics[width=1in,height=1.25in,clip,keepaspectratio]{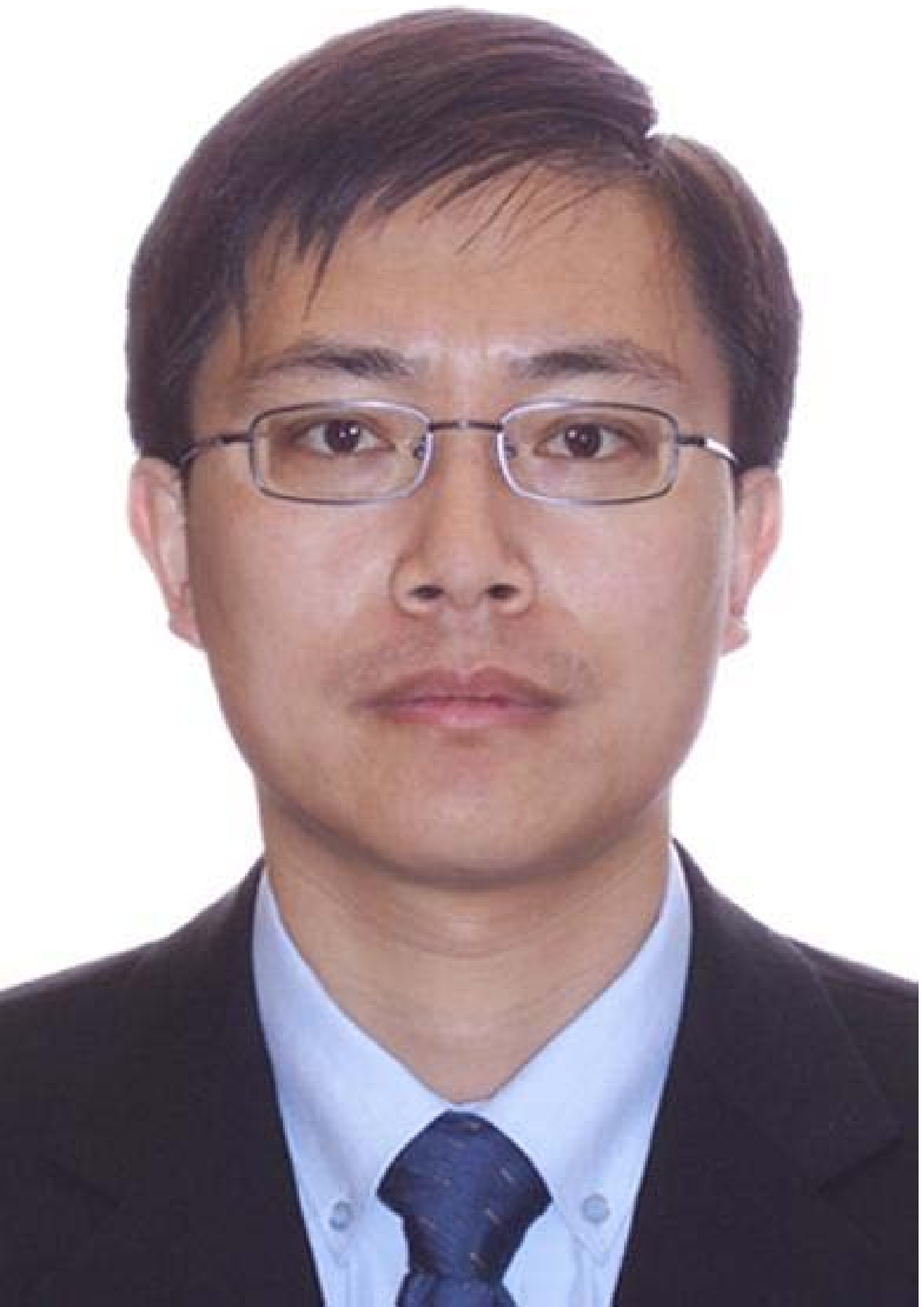}}]{Cheng-Xiang Wang} [S'01-M'05-SM'08] (Cheng-Xiang.Wang@hw.ac.uk) received his Ph.D. degree from Aalborg University, Denmark, in 2004. He joined Heriot-Watt University, U.K., as a lecturer in 2005 and became a professor in August 2011. %He is also an Affiliate Researcher at SNCS Research Center, University of Tabuk, Saudi Arabia.
His research interests include wireless channel modelling and simulation, cognitive radio networks, vehicular communication networks, Large MIMO, green communications, and beyond 4G. He served or is serving as an editor or guest editor for 11 international journals including IEEE Transactions on Vehicular Technology (2011-), IEEE Transactions on Wireless Communications (2007-2009), and IEEE Journal on Selected Areas in Communications. He has edited one book and published one book chapter and over 170 papers in journals and conferences.
\end{IEEEbiography}

\begin{IEEEbiography}[{\includegraphics[width=1in,height=1.25in,clip,keepaspectratio]{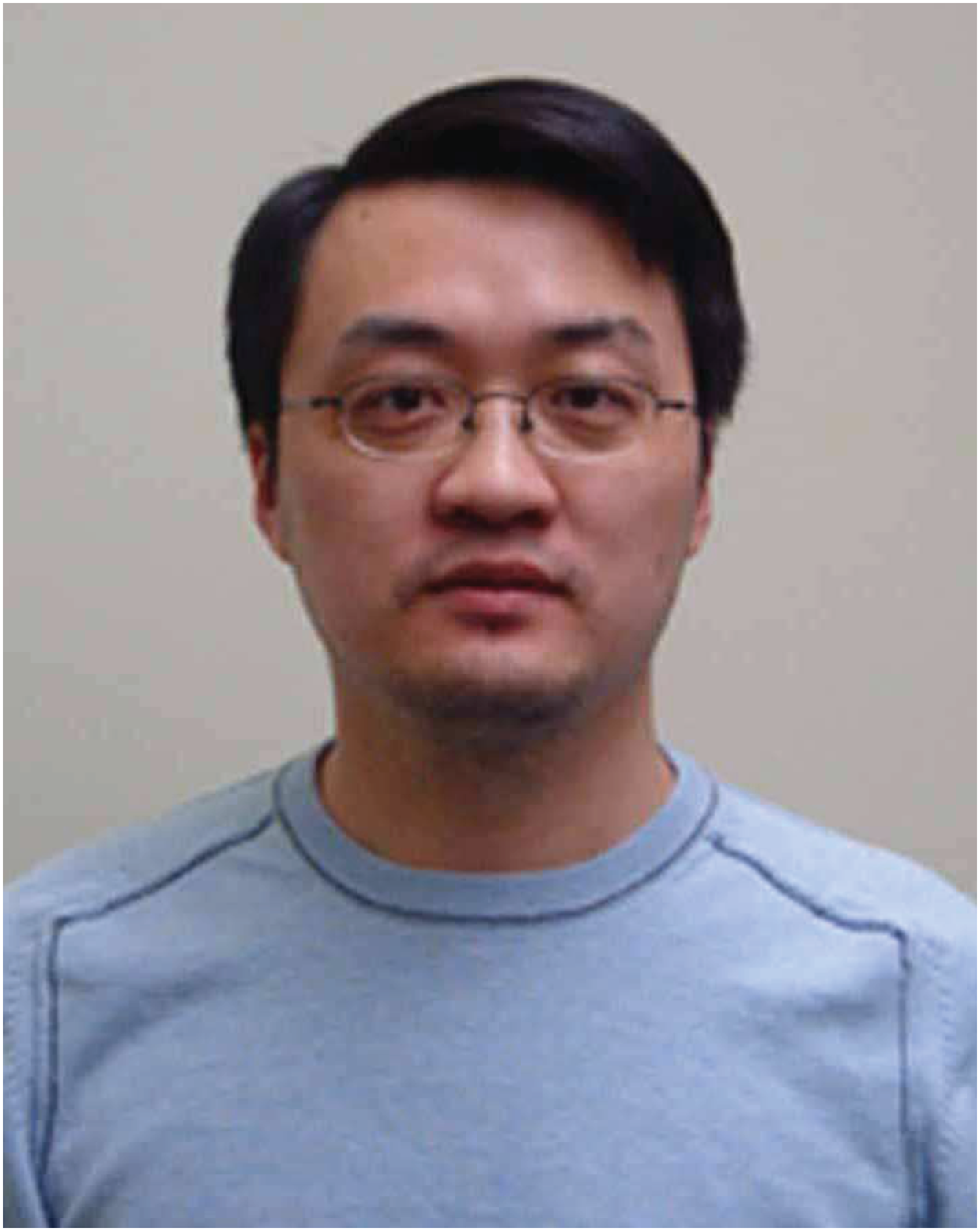}}]{Yunfei Chen} [S'02-M'06-SM'10] (Yunfei.Chen@warwick.ac.uk)
received his B.E. and M.E. degrees in electronics engineering from Shanghai Jiaotong University, Shanghai, P.R.China, in 1998 and 2001, respectively. He received his Ph.D. degree from the University of Alberta in 2006. He is currently working as an Associate Professor at the University of Warwick, U.K. His research interests include wireless communications theory, cognitive radio, and relaying systems.
\end{IEEEbiography}

\newpage

\begin{table}[!h]
\caption{Summary of advantages and disadvantages of narrowband spectrum sensing algorithms.} % title of Table
\begin{tabular}{l | l l} % centered columns (4 columns)
\hline \hline
Spectrum sensing algorithm \;\;& Advantages & Disadvantages \\
\hline %inserts single line
Matched-filtering & Optimal performance & Require prior information \\
 & Low computational cost & of the primary user \\
\hline
Energy detection & Do not require prior information \;\;& Poor performance for low SNR \\
 & Low computational cost & Cannot differentiate users \\
\hline
Cyclostationary feature & Valid in low SNR region & Require partial prior information \\
 & Robust against interference & High computational cost \\
\hline \hline
\end{tabular}
\label{table:compare}
\end{table}

\begin{table}[!ht]
\centering
\tabcolsep 3pt
\caption{Summary of advantages, disadvantages, and challenges of wideband spectrum sensing algorithms.}
\begin{tabular}{p{2.3cm}|p{3cm}|p{3.7cm}|p{3.8cm}|p{4.6cm}}
\hline \hline
Type & \multicolumn{2}{|c|}{Nyquist wideband sensing} & \multicolumn{2}{|c}{Sub-Nyquist wideband sensing} \\
\hline
Algorithm Sub-type & Standard ADC [5]-[6]  & Sweep-tune/filter bank sampling \cite{bank}  & Compressive sensing \cite{scs, cyclic, distributed} & Multi-channel sub-Nyquist sampling \cite{was1, unique1, hongjian} \\ \hline
Advantage & simple structure & low sampling rate, high dynamic range & low sampling rate, signal acquisition cost & low sampling rate, robust to model mismatch \\
 \hline
Disadvantage & high sampling rate, energy cost & high implementation complexity  & sensitive to design imperfections & require multiple sampling channels  \\
 \hline
Challenges & reduce sampling rate, save energy & develop feasible and practical model & improve robustness to design imperfections & relax synchronization requirement\\ \hline \hline
\end{tabular}
\label{table1}
\end{table}

\newpage

\newpage

\newpage

\newpage

\end{document}